# Observations of the nonclassical feature of photon bunching on a beam splitter using coherent photons along the same input port

Sangbae Kim and Byoung S. Ham*
School of Electrical Engineering and Computer Science, Gwangju Institute of Science and Technology
123 Chumdangwagi-ro, Buk-gu, Gwangju 61005, South Korea
(Submitted on Oct. 21, 2021; *bham@gist.ac.kr)

**Abstract**
One of the most striking quantum phenomena is photon bunching resulting from coincidently impinging two-indistinguishable photons on a beam splitter (BS) from two different input ports. Such a nonclassical feature has also been observed even between two independent light sources through either coherence optics resulting in phase locking or post-selected measurements such as quantum beating-based gating. Recently, BS physics regarding quantum features has been discussed using pure coherence optics based on phase basis superposition of the BS. Here, we experimentally demonstrate coherent photon bunching on a BS, where coherent photons come from the same input port. Although the mean values of both output photons are uniform and equal to each other, the mean value of the coincidence measurements between them results in the nonclassical feature of photon bunching at a 50% rate within the classical realm. For this unprecedented result, we discuss the origin of indistinguishability of photons using the wave nature of quantum mechanics to understand the origin of the quantum feature.

**Introduction**
The quantum mechanical interpretation of the two photon-based quantum feature of photon bunching on a beam splitter (BS) involves the indistinguishability between two photons in the space-time domain via quantum superposition between two paths such as in Young's double-slit interference. Such indistinguishability implies coherence between two photons [1], where a two-photon-BS system represents a limited Sorkin's parameter [2] governed by Born's rule [3] regarding the probability amplitudes of a physical quantity in terms of measurements [4-6]. As demonstrated, a single photon-interacting Mach-Zehnder interferometer (MZI) is thus interpreted as a result of Born's rule via path superposition, resulting in an interference fringe [7]. Contrary to the pure quantum mechanical approach using annihilation and creation operators for a single photon with respect to a BS [1], a pure coherence interpretation based on the wave nature of a photon has also been presented recently, exhibiting the same results of photon bunching [8,9]. Here, the same photon bunching phenomenon between two output ports of a BS is discussed using coherent photons along the same input port. According to coherence optics based on time-averaged ensemble measurements, such a photon bunching phenomenon cannot result, and thus the observation of such coherent photon-based quantum features on a BS is beyond the scope of classical optics. Moreover, such a coherence bunching phenomenon has never been observed or discussed.

In a single-photon interacting BS system, a quantum mechanical approach for determining the state of photons in output ports results from path superposition between the probability amplitudes of the photons [1]. In an expanded scheme of a two-photon interacting BS system, the so-called Hong-Ou-Mandel (HOM) effect has been well explained using the same operational approach of quantum mechanics, resulting in the nonclassical feature of photon bunching into either output port [10-15]. Even with independent light source-generated photons, the HOM effect has also been experimentally demonstrated to support the particle nature of quantum mechanics [11,16,17]. On the other hand, such a nonclassical feature on a BS has also been interpreted using pure coherence optics recently, resulting in the presumed $\pm\frac{\pi}{2}$ phase difference between the paired entangled photons [8]. This phase difference between correlated photons is equal to the phase bases of a BS, where the phase bases can be interpreted as pure quantum states for destructive and constructive interference in an interferometric system. Recently, the two-photon intensity correlation has been fully analyzed regarding photon bunching using the wave nature of photons, demonstrating that the origin of indistinguishability resultings from quantum superposition of the phase bases of a photon-BS system [9].

In the present paper, we experimentally demonstrate the quantum feature of photon bunching using coherent photons impinging on a BS from the same input port. Such a coherent photon-based BS scheme with a limited input port is unprecedented, even though the phenomenon of photon bunching is not difficult to guess if independent and incoherent photons are considered as shown in Fig. 1. According to coherence optics, however, coherent photons along the same port impinging on a BS must be treated as one entity, whether it is a single photon or not. In this general coherence approach, such a nonclassical feature of photon bunching cannot be understood as simply due to the same action on a BS. On the other hand, event-by-event measurements in quantum mechanics are timely separated independent processes, where the mean value of individual coincidence detections between output ports should be different from that of the time-averaged products due to the randomness of quantum superposition (discussed below). Thus, our goal in this paper is to bring attention to the fundamental quantum physics of photon bunching on a BS, where the observed coherence photon bunching contradict our common understanding of both quantum mechanics based on the independent photons due to the preset phase coherence and coherence optics due to photon bunching.

**Results**
*Background*

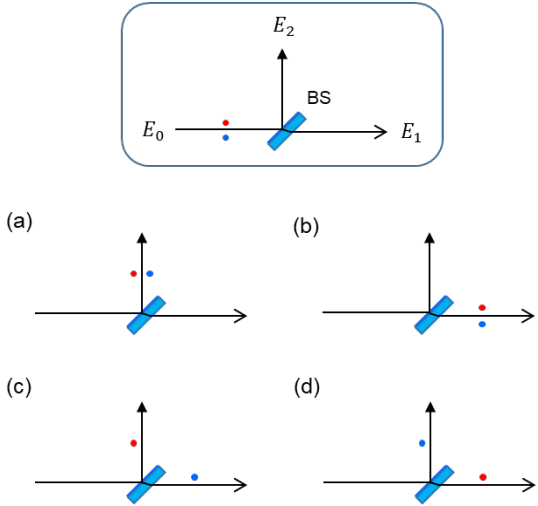

Fig. 1. Schematic of a one-input-two-output beam splitter. (a)-(d) Possible outcomes at equal probability. Inset: input scheme with two coherent photons from the same laser. BS: beam splitter.

The inset of Fig. 1 shows schematic of a coherent photon-based one-input-two-output BS system. The paired input coherent photons (red and blue dots) are generated from a continuous-wave laser via optical attenuation (see Methods). The coincidence measurements between two output photons are individual events-based on quantum mechanics. Thus, the mean value of the measured data is a time-averaged value of all individual events of coincidence measurements. However, this time-averaged mean value based on coincidence measurements is quite different from the classical counterpart of an ensemble average simply due to there being no time-separated events in classical physics. Attenuated laser light is used to satisfy the individual coincidence measurements satisfying the quantum limit with the mean photon number of $\langle n \rangle \ll 1$. The coincidence measurement rules out the majority of vacuum states and thus satisfies sub-Poisson statistics via the inherent post-selection method.

Figures 1(a)-(d) show possible cases of the output photons based on the general concept of quantum mechanics, where the input photons can be considered in general to be independent and incoherent simply due to the nondeterministic phase information. As each particle has two choices in the selection of output port, the two different input photons have four different possible interaction outcomes with the BS as shown. Figures 1(a) and (b) show the quantum feature of photon bunching, while Figs. 1(c) and (d) show the coherence optics with an equal splitting ratio.

One of the most fundamental quantum features is randomness [18], where self-interference is the direct result of randomness via quantum superposition of a single photon itself [7]. For a typical Young's double slit or MZI, the fundamental phase basis set of the system is $\theta \in \{0, \pi\}$, where the phase basis relates to the pure state (basis) of the interferometric system, i.e., constructive and destructive interference. In Fig. 1(a), the fundamental phase basis set of the BS is represented by $\varphi_{BS} \in \left\{-\frac{\pi}{2}, \frac{\pi}{2}\right\}$ resulting from coherence optics [19]. As analyzed previously, satisfying this phase basis for a relative phase between two interacting photons on a BS results in photon bunching [8]. Here, the randomness of the BS phase bases is with respect to the coherent photons in the inset of Fig. 1, where half of the results are for the classical physics of equal splitting ($I_1 = I_2 = I_0$), while the other half of the results are for the quantum feature of photon bunching. Our concern is that the probability amplitude of $E_j$ of each photon for the BS phase bases is the fundamental source of randomness in Born's rule [4-6]. This probability amplitude is a critically different feature of the present analysis compared with both conventional quantum physics and coherence optics.

For each single photon impinging on a BS in the inset of Fig. 1, the photon's amplitude can be denoted as $E_j = E_0 e^{i\varphi_j}$, where $\varphi_j$ is completely random [1]. Due to the coherence optics of the laser cavity, however, the two input photons along the same input port must have the same phase within the given coherence. Thus, the classical approach of coherence optics must show the same results regardless of the number of photons. This thought experiment results in the same conclusion even for the incoherent photon cases in Fig. 1 due to the same input port. Based on the equal probabilities in the output port measurements of the BS, the ensemble product between the output photons must be nonzero according to coherence optics:

$$\begin{bmatrix} E_1 \\ E_2 \end{bmatrix} = \frac{1}{\sqrt{2}} \begin{bmatrix} 1 & \pm i \\ \pm i & 1 \end{bmatrix} \begin{bmatrix} E_0 \\ 0 \end{bmatrix}, \tag{1}$$

where $\pm i$ in equation (1) indicates the two different phase bases of $\varphi_{BS} \in \left\{-\frac{\pi}{2}, \frac{\pi}{2}\right\}$, resulting in a $\pm\frac{\pi}{2}$ phase difference between the two output photons [19]. Thus, the output intensity must be $\langle I_1 \rangle = \langle I_2 \rangle = I_0$ if the same basis combination is applied to the input photons, where $I_0$ is the single photon's intensity.

*Experiments*

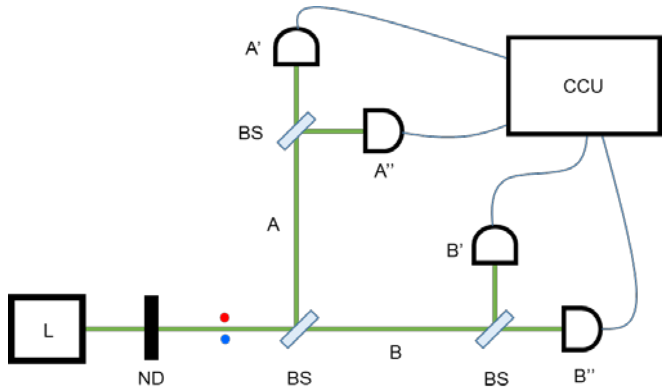

Fig. 2. Schematic of four single photon detector-based coherence measurements for Fig. 1. L: 532 nm laser, ND: neutral density filters, BS: beam splitter, CCU: coincidence counting unit.

Figure 2 shows a modified experimental scheme of Fig. 1. To experimentally confirm the though experiments in Fig. 1, we use coherent photons from an attenuated laser (see Methods). The existence of two coherent photons are verified via coincidence measurements between the output photons. For this, four single photon detectors are used instead of the conventional two detectors. This is to avoid exclusion of photon bunching into either output port from the single photon case, where our detectors cannot distinguish the number of photon. Four single photon detector-based outputs are individually and coincidently measured by a coincidence counting unit (CCU: DE2 of Altera).

Table 1 shows the corresponding results of Fig. 2. The single counts in the second column indicate individual output measurements from each detector. Depending on the mean photon number, the photon statistics results in two or more coincidence photons falling in the measurement time slot (350 ps) of each single photon detector. According to Table 1, the possible four different choices in Fig. 1 have been confirmed at an equal measurement ratio of the mean photon numbers. Thus, the results of Table 1 demonstrate for the coherent photons in Fig. 1, where half of the results are for the nonclassical feature of photon bunching. Half of the rest is for coherence optics-based results at an equal splitting ratio. This nonclassical observation in Table 1 with coherent photons is unprecedented. Thus, careful interpretation regarding the results is necessary because photon bunching cannot be explained nor achieved via coherence optics. Even incoherent photon-based quantum mechanical explanation contradicts the coherent photon source due to violation of phase independent photons. As analyzed below, this is the reason for the mutually exclusive quantum natures between the particle and wave characteristics of a photon [20].

Table 1. Measurements for Fig. 2. Measurement time: 1 sec.

| Port | Single counts | Ports | Coincidence counts | Ports | Triple coincidence counts | $\langle n \rangle$ |
|------|---------------|-------|--------------------|-------|--------------------------|---------------------|
| A'   | 250877.8      | A'A'' | 808.72             | A'A''B' | 2.69                   |                     |
| A''  | 250259.1      | B'B'' | 803.51             | A'A''B'' | 2.25                  |                     |
| B'   | 250441.5      | A'B'  | 798.81             | A'B'B'' | 2.18                   | 0.022               |
| B''  | 250316.4      | A'B'' | 800.58             | A''B'B'' | 2.1                   |                     |
| A'   | 508592.1      | A'A'' | 3440.36            | A'A''B' | 16.17                  |                     |
| A''  | 507361.8      | B'B'' | 3497.19            | A'A''B'' | 16.13                 |                     |
| B'   | 502778.7      | A'B'  | 3483.59            | A'B'B'' | 16.53                  | 0.044               |
| B''  | 504008.3      | A'B'' | 3478.35            | A''B'B'' | 16.36                 |                     |

*Analysis*

Now we analyze the experimental data in Table 1 using the wave nature of quantum mechanics. This analysis is obviously different from conventional coherence optics based on equation (1). Thus, it is a matter of interpreting quantum mechanics based on the wave nature. Because the wave and particle natures of quantum mechanics are mutually exclusive, the photon in the present method cannot be treated as an independent particle either. Instead, coherent photons in the inset of Fig. 1 must be treated as coherent objects impinging on the BS for different phase bases. Because two photons are involved, it is reasonable to think of a superposed photon-BS coupled system, resulting in phase-basis superposition for input photons. This phase-basis superposition of the BS is a unique feature of the present interpretation of photon bunching on a BS, which cannot be explained by conventional coherence (classical) physics or quantum physics based on independent and incoherent photons.

As analyzed in ref. 9, this new quantum interpretation based on the wave nature of a photon results in the following phase basis combinations for two input photons. For this, we define the phase basis of $\varphi_{BS}$ as $[BS]_\pm$, where $[BS]_\pm$ represents a matrix notation in coherence optics [19]:

$$[BS]_\pm = \frac{1}{\sqrt{2}}\begin{bmatrix} 1 & \pm i \\ \pm i & 1 \end{bmatrix}. \tag{2}$$

In equation (2), $[BS]_-$ can be replaced by $[BS]_- = \frac{1}{\sqrt{2}} e^{-i\frac{\pi}{2}} \begin{bmatrix} i & 1 \\ 1 & i \end{bmatrix}$. Here, the global phase of $e^{-i\frac{\pi}{2}}$ has nothing to do with the interferometric result. Thus, there are four combinations based on these phase bases of BS matrices for two input photons, where each coherent photon has a freedom to act independently with the BS matrices. Two of these are for the same basis superposition, while the rest two are for opposite phase-basis superposition.

(i)  For the opposite phase-basis superposition,

$$\begin{bmatrix} E_1 \\ E_2 \end{bmatrix} = \frac{1}{\sqrt{2}} \left\{ \begin{bmatrix} 1 & i \\ i & 1 \end{bmatrix} \pm \begin{bmatrix} 1 & -i \\ -i & 1 \end{bmatrix} \right\} \begin{bmatrix} E_0 \\ 0 \end{bmatrix}$$
$$= \sqrt{2} \begin{bmatrix} 1 & 0 \\ 0 & 1 \end{bmatrix} \begin{bmatrix} E_0 \\ 0 \end{bmatrix} \text{ or } \sqrt{2}i \begin{bmatrix} 0 & 1 \\ 1 & 0 \end{bmatrix} \begin{bmatrix} E_0 \\ 0 \end{bmatrix} \tag{3}$$

In equation (3), the output photons are represented as $E_1 = \sqrt{2} E_0$ and $E_2 = 0$ for the symmetric combination, or $E_1 = 0$ and $E_2 = \sqrt{2} i E_0$ for the antisymmetric combination. In both cases, the coincidence measurements between output photons result in photon bunching into either output port. Here, the representation of the symmetric BS matrix in equation (3) is with respect to the transmitted photon as a reference. If the reference is switched to the reflected photon, then the matrix representation is $[BS]_\pm = \frac{1}{\sqrt{2}} e^{-i\frac{\pi}{2}} \begin{bmatrix} \pm i & 1 \\ 1 & \pm i \end{bmatrix}$. This relation results in the antisymmetric combination in equation (3). The mean values of each output photon are equal because they have the same probability between the symmetric and antisymmetric choices. Thus, equation (3) satisfies the indistinguishable relationship not between the input photons, but between the phase basis choices, resulting in uniform output intensity: $\langle I_1 \rangle = \langle I_2 \rangle = I_0$. In conclusion, the wave nature of a photon does not exclude the resulting quantum feature of photon bunching on a BS.

(ii) For the same phase-basis superposition,

For the remaining two combinations with the same phase-basis superposition applied to both input photons, the following equation is obtained:

$$\begin{bmatrix} E_1 \\ E_2 \end{bmatrix}_\pm = \frac{1}{\sqrt{2}} \begin{bmatrix} 1 & \pm i \\ \pm i & 1 \end{bmatrix} \begin{bmatrix} \sqrt{2} E_0 \\ 0 \end{bmatrix}$$
$$= E_0 \begin{bmatrix} 1 \\ \pm i \end{bmatrix}. \quad (4)$$

From equation (4), the output photons' amplitudes are $E_1 = E_0$ and $E_2 = \pm i E_0$, resulting in $\langle I_1 \rangle = \langle I_2 \rangle = I_0$. This result represents the typical coherence feature with an equal beam splitting ratio. Due to the same probability for the four different outcomes, the experimental coincidence measurements in Table 1 still satisfy classical physics with the lower bound of $g^{(2)}(0) = \frac{1}{2}$ [8, 21].

**Conclusion**

We experimentally demonstrated the nonclassical feature of photon bunching using coherent photons simultaneously impinging on a BS along the same input port. Unlike the common understanding of coherence optics resulting in an equal splitting ratio, the nonclassical feature was analyzed using the phase-basis combinations of the photon-BS system. The concept of indistinguishability in this quantum features was understood experimentally, where this indistinguishability was not from individual photons but from a random phase basis choice of the BS, resulting in quantum superposition of the interacting photons. Moreover, photon bunching indicated that the quantum process could be treated as a special action of instantaneous measurement via coincidence detection between the paired output photons. Although coherence optics prohibits such a nonclassical feature due to the measurement method of ensemble averages, instantaneous photon detection between two output ports results in photon bunching at a 50% rate. Thus, the wave nature of quantum mechanics describes the same quantum feature in a different way using the phase basis of an interferometric system. This conclusion implies that there is no fundamental difference between quantum and classical physics in an interferometric system according to Born's rule tested over decades [4-6], where this quantum feature is not due to fundamental physics but instead due to different measurement techniques.

**Acknowledgment**
BSH acknowledges that this work was motivated by Prof. J. Lee of Hanyang University, S. Korea for the discussion of coherent photon interaction with a BS.

**Methods**
    1. Dark count measurements for a single photon detection module

The dark count rate of the single photon counting module (Excelitas AQRH-SPCM-15) is measured at $27 \pm 5$ (counts/s) in an experimental dark-room condition.
    2. Attenuated 532 nm laser

A Coherent cw laser (Verdi V10) is stabilized at ~1 % in the output intensity variation for the fixed output power of 10 mW. For the 2 M cps rate in Table 2, optical density filters with OD 13 is used, resulting in $1~\text{fW}$ power and a mean photon number of $\langle n \rangle \sim 0.04$.
    3. Coincidence detection

The SPCM-generated electrical pulses are sent to the four-channel coincidence counting module (CCU: DE2 FPGA, Altera). The pulse duration of each electrical signal of a single photon detection by the SPCM is ~10 ns. Both single and coincidence counting numbers are simultaneously measured by the CCU for 1 s acquisition time for Table 1.

**Data availability**
Data for figures and Table are available upon reasonable requests.

**Funding**
This work was supported by GIST-GRI 2021 and GIST-GTI 2021 via Practical Research and Development support program and ICT R&D program of MSIT/IITP (2021-0-01810), development of elemental technologies for ultra-secure quantum internet.

**Author contributions**

B.S.H. conceived the idea, developed the theory, analyzed the experimental data, and wrote the paper. SK conducted the experiments.

**Competing interests**
The authors declare no competing interests.